\documentclass[journal]{IEEEtran}
		
\newcommand{\E}{\mathrm{E}}

\usepackage{epsf,psfrag,amssymb,amsfonts,color,cite,fancybox}
\usepackage[mathscr]{eucal}
\usepackage[dvips]{graphicx}
\usepackage{caption}
\usepackage{subcaption}
\usepackage{ifpdf}
\usepackage{longtable}	
\usepackage{multicol}
\usepackage{float}
\usepackage{epstopdf}
\usepackage{bm}
\usepackage{multirow}
\usepackage[scientific-notation=true]{siunitx}

\ifCLASSINFOpdf

\else

\fi

\usepackage[cmex10]{amsmath}
\begin{document}


\title{Estimating Dynamic Load Parameters from Ambient PMU Measurements}

\author{Xiaozhe Wang,~\IEEEmembership{Member,~IEEE.}
\thanks{Xiaozhe Wang is with the Department of Electrical and Computer Engineering, McGill University, Montreal, QC H3A 0G4, Canada. Email: xiaozhe.wang2@mcgill.ca.}
}

\maketitle

\begin{abstract}
In this paper, a novel method to estimate dynamic load parameters via ambient PMU measurements is proposed. Unlike conventional parameter identification methods, the proposed algorithm does not require the existence of large disturbance to power systems, and is able to provide up-to-date dynamic load parameters consistently and continuously. The accuracy and robustness of the method are demonstrated through numerical simulations.
\end{abstract}


\IEEEpeerreviewmaketitle

\section{Introduction}
Load modelling and identification are of great importance to the security and stability of power systems. While the accurate models are available for generators, transmission lines and control devices, load modeling is still a challenging and open subject due to the fact that electric load at each substation is an aggregation of numerous individual loads with different behaviors \cite{Hill:1994}-\cite{Hiskens:2006}. In addition, the poor measurements, modeling, exchange information, as well as the uncertainties in customers behaviors/devices further result in load uncertainties \cite{Turistsyn:2016}. Indeed, load uncertainty is one of the main factors that affect the accuracy of the power dynamic models implemented by system operators over the world\cite{Hiskens:2006}.

Generally speaking, the load uncertainty comes from both model structure and parameter values. It has been shown in previous literature \cite{Najafabadi:2012}-\cite{Kao:1994} that the use of different load models leads to different and even contradictory results for dynamic stability studies. Even though the applied model structure is verified, different parameter values may also yield different damping performances in small signal stability \cite{Turistsyn:2016}\cite{Hiskens:1995}\cite{Hiskens:1995_1}. For instance, different time constants of loads may lead to either asymptotically stable system or systems experiencing oscillations (i.e., Hopf bifurcation occurs)\cite{Turistsyn:2016}. Both load modelling and parameter identifications are essential in studying the dynamic behaviors of power systems. This paper mainly focuses on parameter identification for a generic dynamic load model that is suitable for small signal stability analysis.

Different methods for dynamic load parameter identification have been proposed, which can be classified into two categories: component-based approach \cite{Price:1988} and measurement-based approach \cite{Chiang:2006}-\cite{Wen:2003}. The latter approach is more commonly applied because real-time load variations and dynamic characteristics can be taken into account\cite{Ajjarapu:2009}. Measurement-based model identification is typically solved through optimization methods that minimize the error between the measured output variables and the simulated ones. In particular, the nonlinear least-square curve fitting method has been implemented in \cite{Hill:1994}\cite{Chiang:2006}-\cite{Hill:2008}. Genetic algorithms, neural network-based methods and other heuristic techniques have been applied in \cite{Karlsson:1996}-\cite{Wen:2003}. However, the optimization-based methods are time consuming and thus can not be implemented online\cite{Najafabadi:2012}. More importantly, all those methods require measurement data from dynamic behaviors of system under big disturbances (e.g., during faults), which is not always available \cite{Han:2009}. Indeed, the variation of load parameters may be much faster than the occurrence rate of natural disturbances \cite{Hill:1994}. 

In this paper, we propose a novel measurement-based method for dynamic load identification in ambient conditions, which does not require the existence of large disturbance. Particularly, the method combines the statistical properties extracted from PMU measurements and the inherent model knowledge, and is able to provide fairly accurate estimations for parameter values in near real-time. Note that a generic dynamic load model is implemented in this paper which is suitable for the purposes of small signal stability analysis and damping performance \cite{Turistsyn:2016}\cite{Hiskens:1995}\cite{Hiskens:1995_1}\cite{Turistsyn:2015}. The proposed method can be implemented in online security analysis to provide up-to-date dynamic load parameters accurately. 


The rest of the paper is organized as follows. Section \ref{sectionmodel} introduces the power system stochastic dynamic model. Particularly, the generic dynamic load model used in small signal stability is presented. Section \ref{sectionmethod} elaborates the proposed method for estimating parameters of dynamic loads. Section \ref{sectioncasestudy} presents the validation of the proposed method through numerical simulations. The impact of measurement noise is also investigated. Conclusions and perspectives are given in Section \ref{sectionconclusion}.

\section{power system stochastic dynamic model}\label{sectionmodel}
Although we focus on load models, generator models are also incorporated to provide more realistic simulations. Specifically, the classical generator model which can reasonably represent the dynamics of generator in ambient conditions is implemented. The power system buses are numbered as follows: load buses: $k = 1,2,...,m$, and generators: $i=m+1,...,N$. Particularly, to include the effects of the loads, the structure preserving model \cite{Chiang:book}\cite{Pai:2012} is used:

\small{
\begin{eqnarray}
\dot{\delta}_i&=&{\omega_i}\label{swing-1}\\
M_i\dot{{\omega_i}}&=&{P_{mi}}-{P_{Gi}}(\delta_i,\theta_i,V_i)-{D_i}{\omega_i}\label{swing-2}\\ 
P_{Gi}(\delta_i,\theta_i,V_i)&=&\sum_{k=1}^N|V_i||V_k|(G_{ik}\cos\theta_{ik}+B_{ik}\sin\theta_{ik})\\
Q_{Gi}(\delta_i,\theta_i,V_i)&=&\sum_{k=1}^N|V_i||V_k|(G_{ik}\sin\theta_{ik}-B_{ik}\cos\theta_{ik})
\end{eqnarray}}
\normalsize
where
\begin{table}[!ht]\normalsize
\begin{tabular}{ll}
  $\delta_i$& generator rotor angle\\
  $\omega_i$& generator angular frequency\\
  $M_i$& inertial constant \\
  $P_{mi}$ & mechanical power input \\
  $P_{Gi}(\delta_i,\theta_i,V_i)$ & real power injection \\
  $Q_{Gi}(\delta_i,\theta_i,V_i)$ & reactive power injection \\
  $D_i$ & damping coefficient \\
  $N$& total number of buses\\
  $\theta_{ij}$& voltage angle difference between bus $i$ and $j$\\
  $|V_i|$ & voltage magnitude\\
  $G_{ij}$& line conductance between bus $i$ and $j$\\
  $B_{ij}$& line susceptance between bus $i$ and $j$
  \end{tabular}
\end{table}

\noindent The detailed expressions of $P_{Gi}(\delta_i,\theta_i,V_i)$ and $Q_{Gi}(\delta_i,\theta_i,V_i)$ are neglected here for simplicity and can be found in many books (e.g., \cite{Chiang:book}).
\normalsize

Regarding dynamic loads, we use the following first-order load model proposed in \cite{Turistsyn:2015} that can represent the common types of loads (e.g., induction motors, thermostatically controlled loads) in ambient conditions:
\small{
\begin{eqnarray}
\dot{g}_k&=&-\frac{1}{\tau_{g_k}}(P_k-P_{k}^s)\label{loaddynamicp}\\
\dot{b}_k&=&-\frac{1}{\tau_{b_k}}(Q_k-Q_{k}^s)\label{loaddynamicq}\\
P_k=g_kV_k^2&=&\sum_{j=1}^N|V_k||V_j|(-G_{kj}\cos\theta_{kj}-B_{kj}\sin\theta_{kj})\\
Q_k=b_kV_k^2&=&\sum_{j=1}^N|V_k||V_j|(-G_{kj}\sin\theta_{kj}+B_{kj}\cos\theta_{kj})
\end{eqnarray}}
\normalsize
where
\begin{table}[!ht]\normalsize
\begin{tabular}{ll}
  $g_k$& effective conductance of the load\\
  $b_k$& effective susceptance of the load \\
  $\tau_{gk}$ & active power time constant of the load \\
  $\tau_{bk}$ & reactive power time constant of the load \\
  $P_k$ &   real power demand of the load\\
  $Q_k$ &   reactive power demand of the load\\
  $P_k^s$ & steady-state real power demand of the load \\
  $Q_k^s$ & steady-state reactive power demand of the load \\
\end{tabular}
\end{table}

\noindent The values $P_k^s$ and $Q_k^s$ describe the static (steady-state) power characteristics of the loads achieved in equilibrium. The instant real power and reactive power consumption can be characterized by the effective conductance $P_k=g_kV_k^2$ and susceptance $Q_k=b_kV_k^2$ at any time. The time constants $\tau_{gk}$ and $\tau_{bk}$ that typically depend on voltage and frequency represent the instant relaxation rate of the load.

To incorporate load variation, we apply a similar approach used in \cite{Nwankpa:2000}\cite{Nwankpa:1992} and modify the set of load equations (\ref{loaddynamicp})-(\ref{loaddynamicq}) as follows:
\begin{eqnarray}
\dot{g}_k&=&-\frac{1}{\tau_{g_k}}[P_k-P_{k}^s(1+\sigma_k^p\xi_k^p)]\label{stoloaddynamicp}\\
\dot{b}_k&=&-\frac{1}{\tau_{b_k}}[Q_k-Q_{k}^s(1+\sigma_k^q\xi_k^q)]\label{stoloaddynamicq}
\end{eqnarray}
where the steady-state real and reactive load demands are perturbed with independent Gaussian noise from their initial values. Specifically, $\xi_k^p$ and $\xi_k^q$ are standard Gaussian noise, and $\sigma_k^p$ and $\sigma_k^q$ represent the noise intensities for static real and reactive power, respectively.

As discussed in \cite{Turistsyn:2016}\cite{Turistsyn:2015}, this dynamic load model can naturally represent the most common types of loads in ambient conditions such as thermostatic load, induction motor, power electronic converter, aggregate effects of distribution load tap changer (LTC) transformers, etc. 
However, the range of time constants is considerably large ranging from cycles to several minutes, and even hours for different types of loads. For industrial plants, such as aluminum smelters, the time constants are in the range of $0.1$s to $0.5$s; for tap changers and other control devices, they are in the range of minutes; for heating load, they may range up to hours \cite{Hiskens:1995_1}. As a result, the uncertainty of composition of different types of loads can be aggregated in time constants $\tau_{g}$ and $\tau_{b}$\cite{Turistsyn:2016}. This is reasonable in the situations when the network characteristics are known, generator models are validated and static load characteristics are understood better than their dynamic response which is the case in practical situations. In addition to a wide range of time constants, the variation of $\tau_g$ and $\tau_b$ can also be fast. For example, $\tau_b$ may change from $0.1$s to $24.1$s in one day (see Table I, II in \cite{Hill:1994}).

Because of wide range and fast variation of time constants $\tau_g$ and $\tau_b$, they need to be updated frequently to ensure the accuracy of dynamic load models used in online security and stability analysis. Conventionally, $\tau_g$ and $\tau_b$ are estimated from dynamic data by perturbing the system, for example, through changing the transformer tap\cite{Cutsem:book}. However it is impractical to perturb the system frequently for estimating parameter values of loads. In this paper, we propose a novel method to estimate $\tau_g$ and $\tau_b$ for the loads of interests from ambient PMU measurements in daily operation. In particular, the estimation process does not require the existence of disturbance to the system.


\section{Methodology}\label{sectionmethod}
In ambient conditions, the stochastic dynamic load equations (\ref{stoloaddynamicp})-(\ref{stoloaddynamicq}) can be linearized as below:
\begin{eqnarray}
\left[\begin{array}{c}\dot{\bm{g}}\\\dot{\bm{b}}\end{array}\right]&=&\left[\begin{array}{cc}-{T_g}^{-1}\frac{\partial{\bm{P}}}{\partial{\bm{g}}}&\bm{0}\\
\bm{0}&-{T_b}^{-1}\frac{\partial{\bm{Q}}}{\partial{\bm{b}}}\end{array}\right]
\left[\begin{array}{c}{\bm{g}}\\{\bm{b}}\end{array}\right]\nonumber\\
&+&\left[\begin{array}{cc}{T_g}^{-1}P^s\Sigma^p&\bm{0}\\
\bm{0}&{T_b}^{-1}Q^s\Sigma^q\end{array}\right]\nonumber
\left[\begin{array}{c}{\bm{\xi^p}}\\{\bm{\xi^q}}\end{array}\right]\\
&=&A\left[\begin{array}{c}{\bm{g}}\\{\bm{b}}\end{array}\right]
+B\left[\begin{array}{c}{\bm{\xi^p}}\\{\bm{\xi^q}}\end{array}\right]
\end{eqnarray}
where
\begin{table}[!ht]\normalsize
\begin{tabular}{ll}
$\bm{g}=[g_1,...,g_m]^T$, & $\bm{b}=[b_1,...,b_m]^T$, \\
$T_g=\mbox{diag}[\tau_{g1},...,\tau_{gm}]$, &$T_b=\mbox{diag}[\tau_{b1},...,\tau_{bm}]$,\\ $\bm{P}=[P_1,...,P_m]^T$,& $\bm{Q}=[Q_1,...,Q_m]^T$,\\
$P^s=\mbox{diag}[P_1^s,...,P_m^s]$,& $Q^s=\mbox{diag}[Q_1^s,...,Q_m^s]$, \\ $\Sigma^p=\mbox{diag}[\sigma^p_1,...,\sigma^p_m]$,& $\Sigma^q=\mbox{diag}[\sigma^q_1,...,\sigma^q_m]$, \\ $\bm{\xi^p}=[\xi^p_1,...,\xi^p_m]^T$,& $\bm{\xi^q}=[\xi^q_1,...,\xi^q_m]^T$.
\end{tabular}
\end{table}

It is observed that $[\bm{g},\bm{b}]^T$ is a vector Ornstein-Uhlenbeck process that is stationary, Gaussian and Markovian \cite{Gardiner:2009}\cite{Wangxz:2015}. Particularly, if the state matrix $A$ is stable, the stationary covariance matrix $C_{\bm{xx}}=\left[\begin{array}{cc}C_{\bm{gg}}&C_{\bm{gb}}\\C_{\bm{bg}}&C_{\bm{bb}}\end{array}\right]$  can be shown to satisfy
the following Lyapunov equation\cite{Gardiner:2009}\cite{Hines:2015}:
\begin{equation}
  AC_{\bm{xx}}+C_{\bm{xx}}A^T=-BB^T \label{lyapunov}
\end{equation}
which nicely combines the model knowledge and the statistical properties of state variables.

Since $P_k=g_kV_k^2$, we have:
\begin{equation}
\frac{\partial{P_k}}{\partial{g_j}}=\left\{\begin{array}{cc}V_k^2+2g_jV_k\frac{\partial{V_k}}{\partial{g_j}}\approx{V_k^2}&\mbox{if }j=k\\
2g_jV_k\frac{\partial{V_k}}{\partial{g_j}}\approx 0&\mbox{if }j\not=k\end{array}\right.
\end{equation}
under the assumption that $\triangle V_k\approx 0$ in ambient conditions. Similar relation can be obtained for $\frac{\partial{Q_k}}{\partial{b_j}}$. As a result, the Jacobian matrix $A$ satisfies:
\begin{eqnarray}
A\approx \left[\begin{array}{cc}-{T_g}^{-1}V^2&\bm{0}\\
\bm{0}&-{T_b}^{-1}V^2\end{array}\right]\label{approxA}
\end{eqnarray}
where $V=\mbox{diag}[V_1,...,V_m]$. Substituting (\ref{approxA}) and the detailed expression of $C_{\bm{xx}}$ and $B$ into (\ref{lyapunov}),
and performing algebraic simplification, we have:
\begin{eqnarray}
&&C_{\bm{gg}}=\frac{1}{2}T_g^{-1}(P^s)^2(\Sigma^p)^2V^{-2}\label{Cgg}\\
&&C_{\bm{bb}}=\frac{1}{2}T_b^{-1}(Q^s)^2(\Sigma^q)^2V^{-2}\label{Cbb}\\
&&C_{\bm{gb}}=C_{\bm{bg}}=0
\end{eqnarray}
Particularly, we utilize the relations (\ref{Cgg})-(\ref{Cbb}) that link the measurements of stochastic load variation to the physical model, and provide an ingenious way to estimate the dynamic parameters $T_g$ and $T_b$ from measurements.

In practical applications, $V$, $C_{\bm{gg}}$, and $C_{\bm{bb}}$ need to be acquired or estimated from limited PMU measurements. A window size of $1000$s is used in the examples of this paper where time constants are up to several seconds. Note that the larger the time constants, the longer the sample window is needed to ensure accuracy. First, the sample mean $\bar{V}$ can be used as an estimation of $V$, then $\bm{g}$ and $\bm{b}$ can be estimated from PMU measurements (i.e., phasors $V_k$ and $I_k$) as follows:
\begin{eqnarray}
g_k&=&\mbox{Re}\{\frac{I_k}{V_k}\}\label{g}\\
b_k&=&\mbox{Im}\{\frac{I_k}{V_k}\}\label{b}
\end{eqnarray}
Regarding the covariance matrix $C_{\bm{g}\bm{g}}=\E[(\bm{g}-\E[\bm{g}])(\bm{g}-\E[\bm{g}])^T]$ and $C_{\bm{b}\bm{b}}=\E[(\bm{b}-\E[\bm{b}])(\bm{b}-E[\bm{b}])^T]$, we use their unbiased estimators---sample covariance matrixes $Q_{\bm{g}\bm{g}}$ and $Q_{\bm{b}\bm{b}}$ in practice, each entry of which is calculated as below:
\begin{eqnarray}
Q_{g_kg_j}&=&\frac{1}{n-1}\sum_{i=1}^n(g_{k}(i)-\bar{g}_k)(g_{j}(i)-\bar{g}_j)\label{qgg}\\
Q_{b_kb_j}&=&\frac{1}{n-1}\sum_{i=1}^n(b_{k}(i)-\bar{b}_k)(b_{j}(i)-\bar{b}_j)\label{qbb}
\end{eqnarray}
where $\bar{{g}}_k$ and $\bar{{b}}_k$ denote the sample mean of ${g_k}$ and ${b_k}$, respectively, and $n$ is the sample size.

Therefore, the proposed algorithm can be summarized as follows. We assume that PMUs are installed at the substations that the (aggregated) loads of interests are connected to. We also assume that the static characteristics of loads are well understood such that $P^s$, $Q^s$, $\Sigma^p$ and $\Sigma^q$ are prior known, which is reasonable as shown in\cite{Nwankpa:2000}\cite{Nwankpa:1992}. Then the following algorithm provides an estimation of $T_g$ and $T_b$ for the dynamic loads from ambient PMU measurements:

\begin{description}[\IEEEusemathlabelsep\IEEEsetlabelwidth{Step 1.}]
\item [\textbf{{Step 1.}}] Compute the sample mean $\bar{V}$ and estimate $\bm{g}$ and $\bm{b}$ from PMU measurements by (\ref{g})-(\ref{b}).
\item [\textbf{{Step 2.}}] Calculate the sample covariance matrix $Q_{\bm{g}\bm{g}}$ and $Q_{\bm{b}\bm{b}}$ by (\ref{qgg})-(\ref{qbb}).
\item [\textbf{{Step 3.}}] Approximate $T_g$ and $T_b$ as blow:
\begin{eqnarray}
T_g&=&\frac{1}{2}(P^s)^2(\Sigma^p)^2\bar{V}^{-2}Q_{\bm{gg}}^{-1}\label{Tg}\\
T_b&=&\frac{1}{2}(Q^s)^2(\Sigma^q)^2\bar{V}^{-2}Q_{\bm{bb}}^{-1}\label{Tb}
\end{eqnarray}
\end{description}
Note that (\ref{Tg})-(\ref{Tb}) are acquired by a simple algebraic manipulation of (\ref{Cgg})-(\ref{Cbb}).

\section{case studies}\label{sectioncasestudy}

In this section, the proposed algorithm to estimate time constants of dynamic loads are validated through numerical simulations. Furthermore, the robustness of the proposed method to measurement noise is also demonstrated via simulation. All case studies were done in PSAT-2.1.9 \cite{Milano:PSAT}.

\subsection{Validation of the Method}\label{sectionvalidation}

\begin{figure}[!ht]
\centering
\begin{subfigure}[t]{0.52\linewidth}
\includegraphics[width=1.8in ,keepaspectratio=true,angle=0]{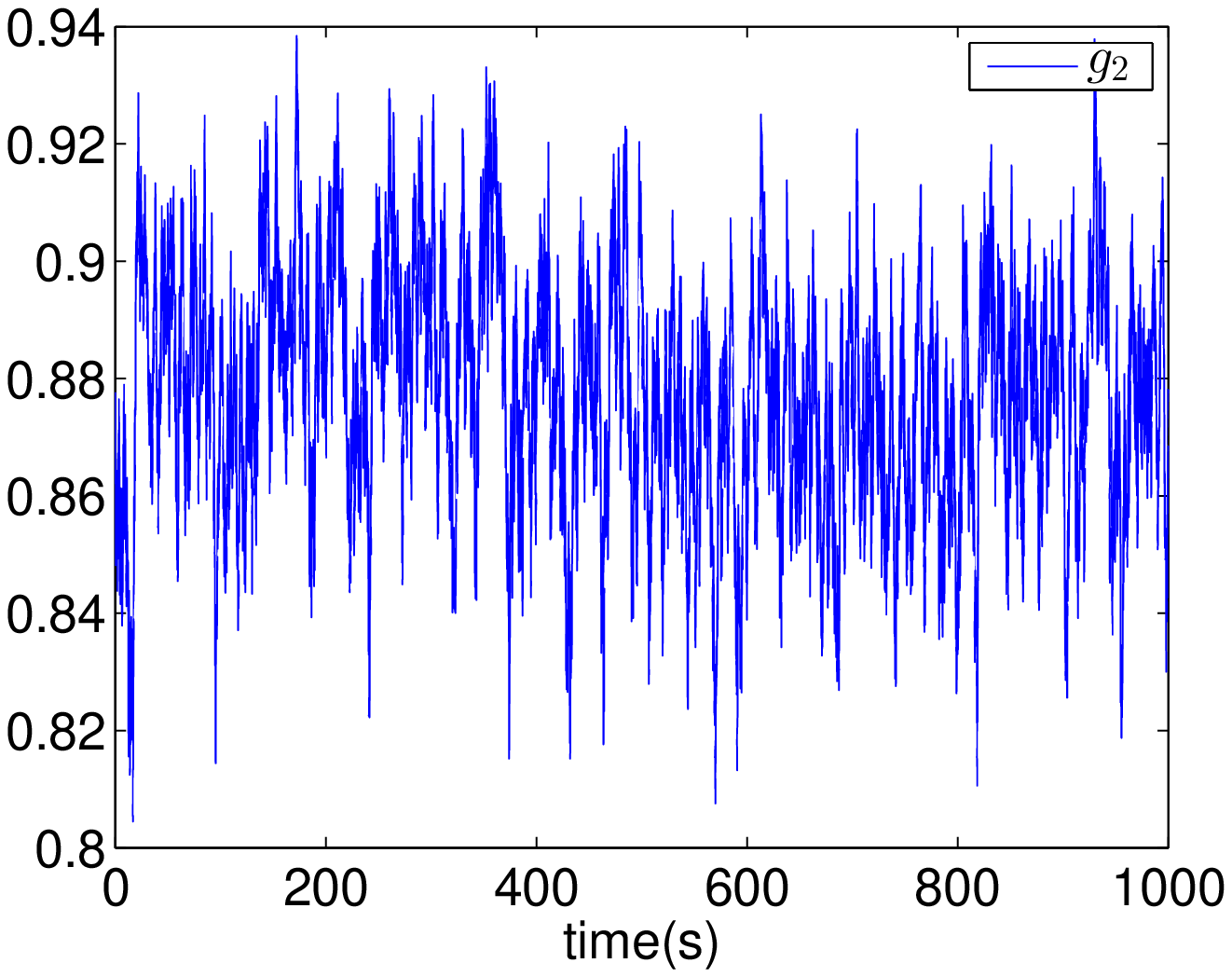}
\caption{Trajectory of $g_2$ on [0s,1000s]}\label{g2}
\end{subfigure}%
\begin{subfigure}[t]{0.48\linewidth}
\includegraphics[width=1.8in ,keepaspectratio=true,angle=0]{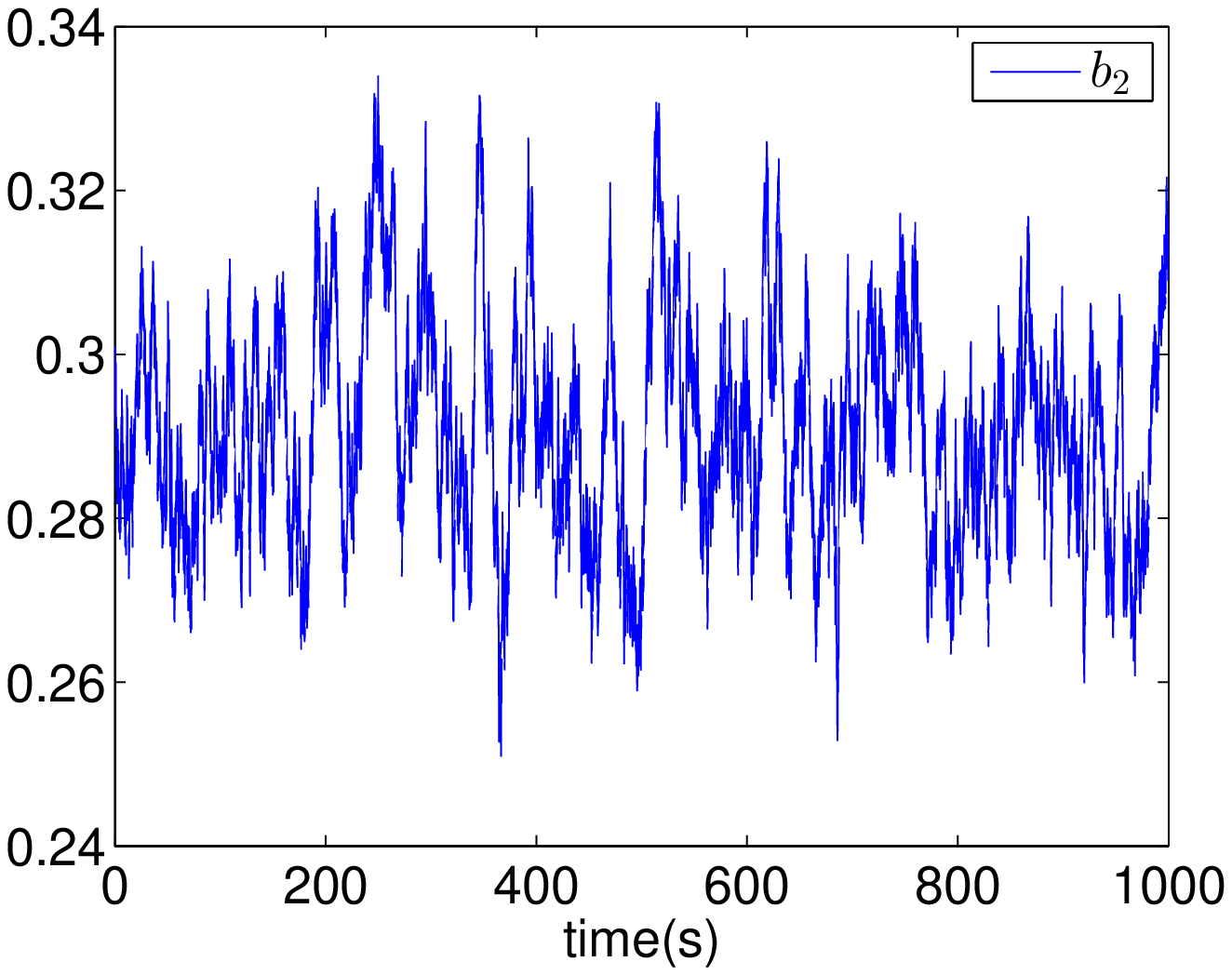}
\caption{Trajectory of $b_2$ on [0s,1000s]}\label{b2}
\end{subfigure}
\begin{subfigure}[t]{0.5\linewidth}
\includegraphics[width=1.8in ,keepaspectratio=true,angle=0]{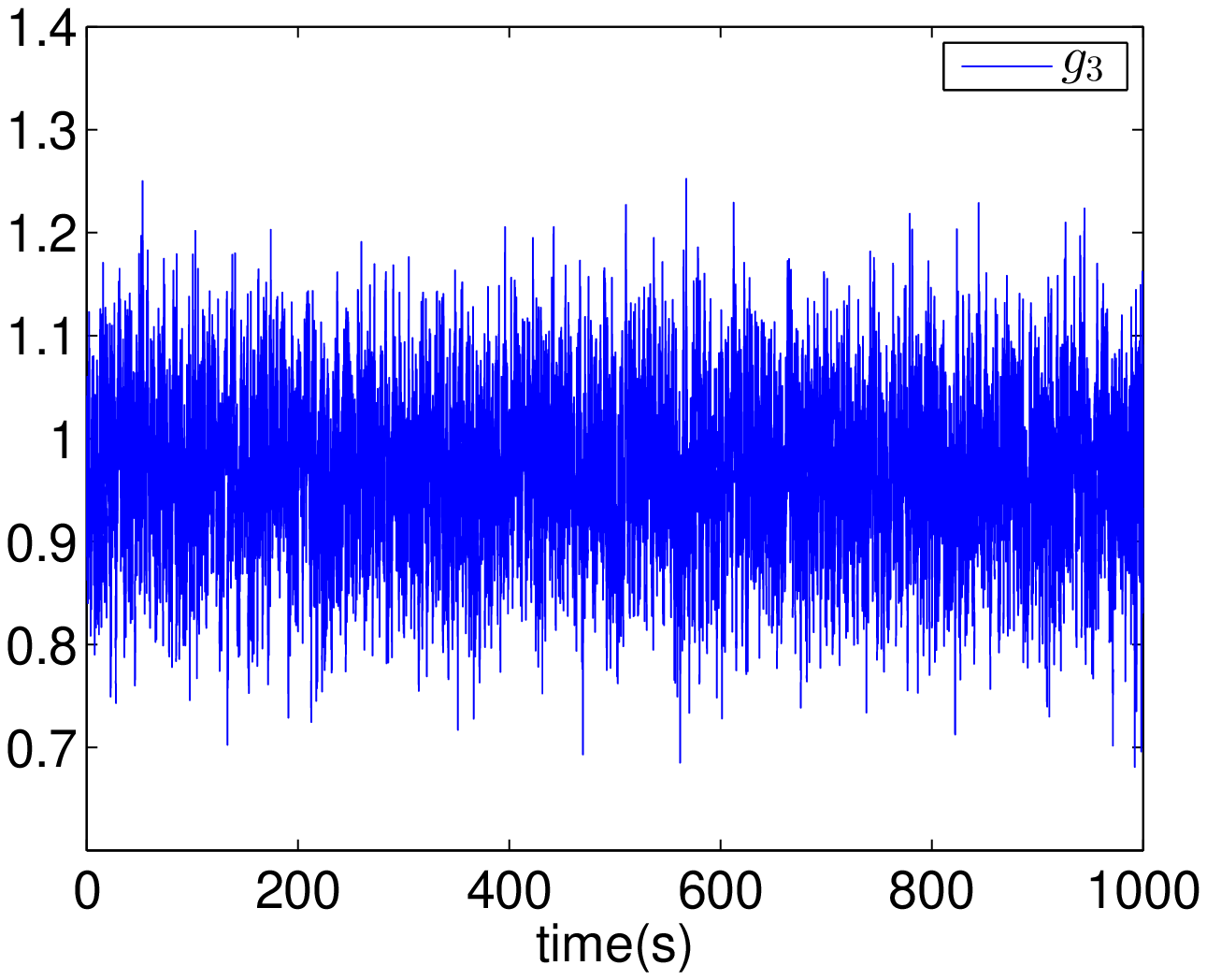}
\caption{Trajectory of $g_3$ on [0s,1000s]}\label{g3}
\end{subfigure}%
\begin{subfigure}[t]{0.5\linewidth}
\includegraphics[width=1.8in ,keepaspectratio=true,angle=0]{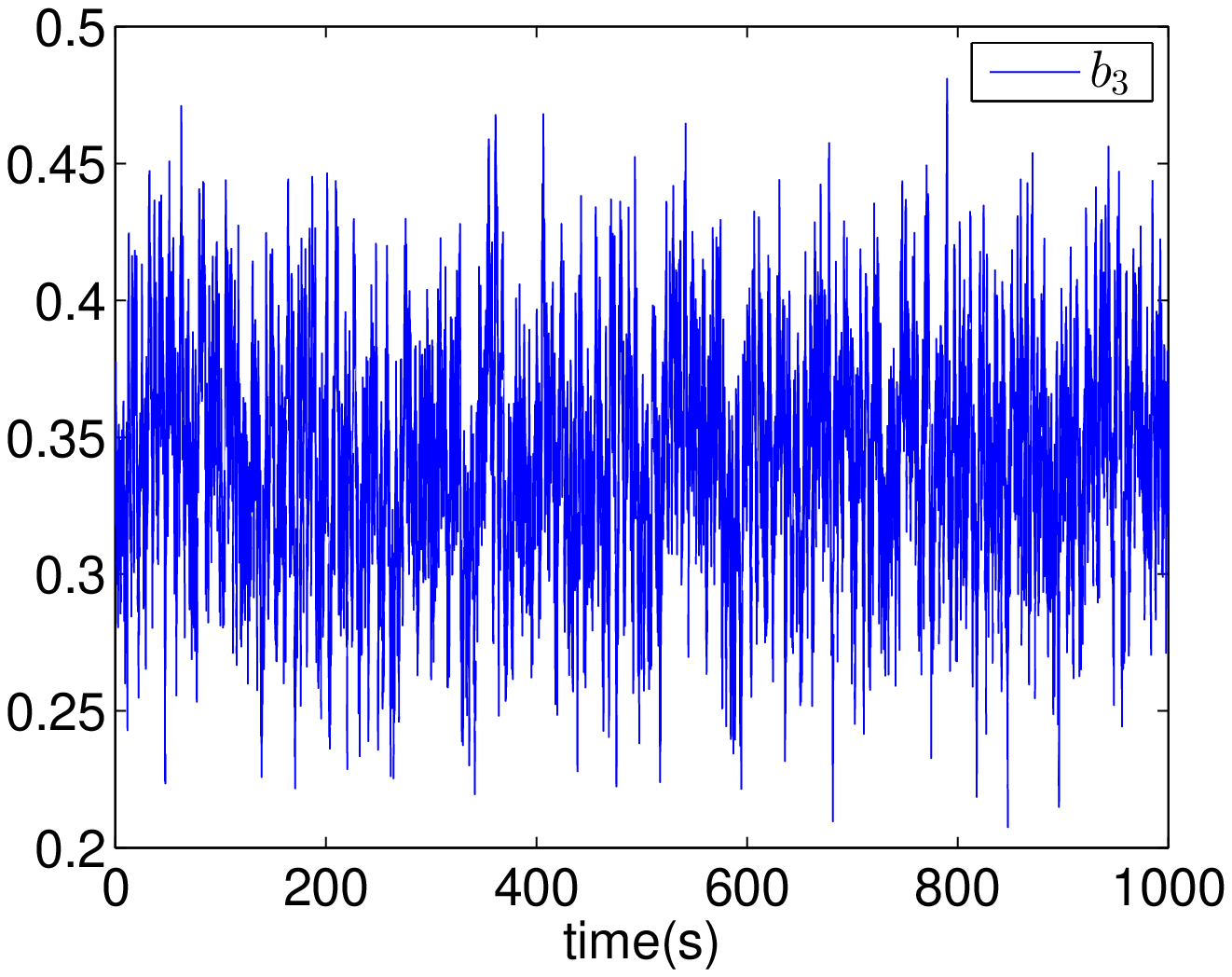}
\caption{Trajectory of $b_3$ on [0s,1000s]}\label{b3}
\end{subfigure}
\begin{subfigure}[t]{0.5\linewidth}
\includegraphics[width=1.8in ,keepaspectratio=true,angle=0]{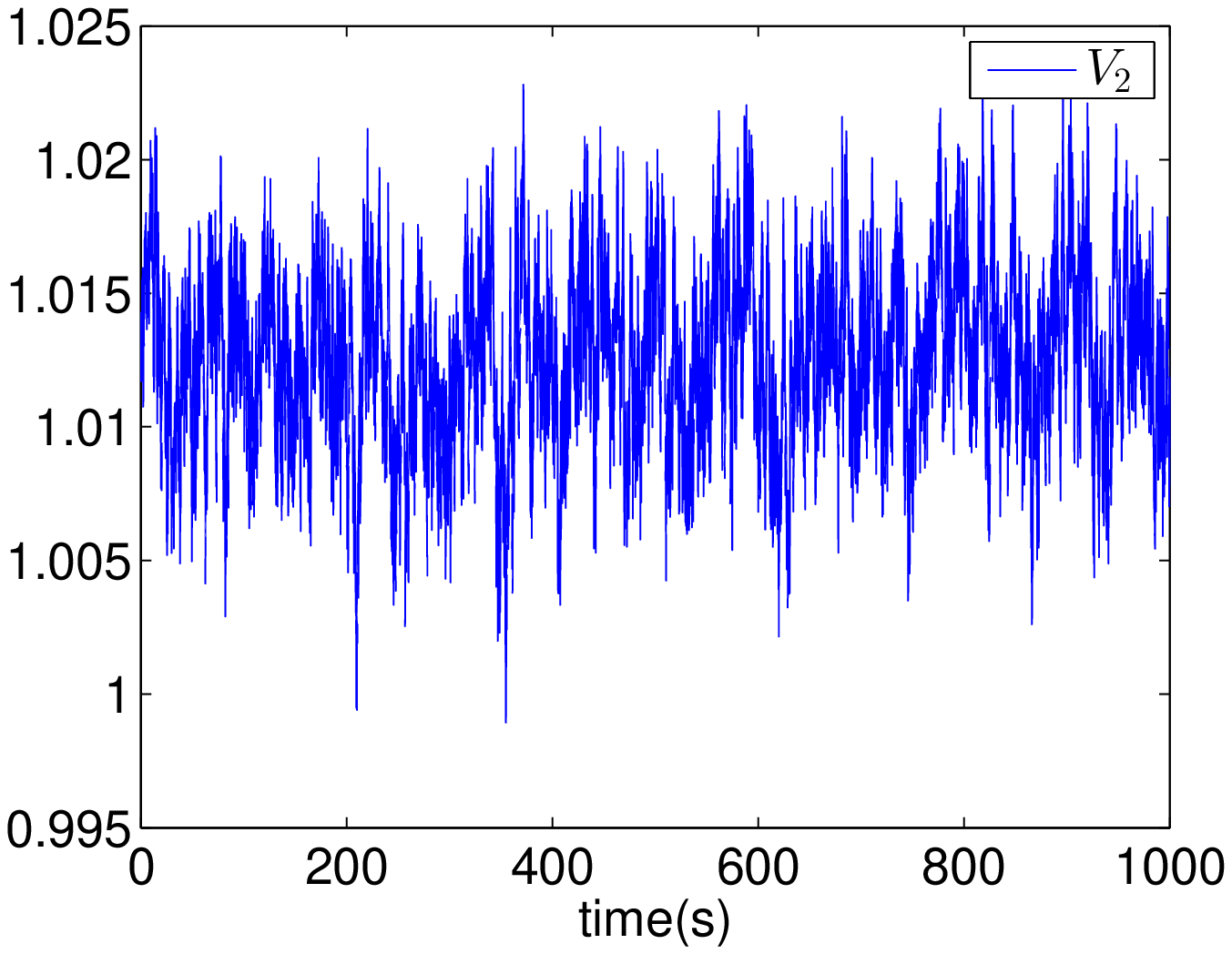}
\caption{Trajectory of $V_2$ on [0s,1000s]}\label{g3}
\end{subfigure}%
\begin{subfigure}[t]{0.5\linewidth}
\includegraphics[width=1.8in ,keepaspectratio=true,angle=0]{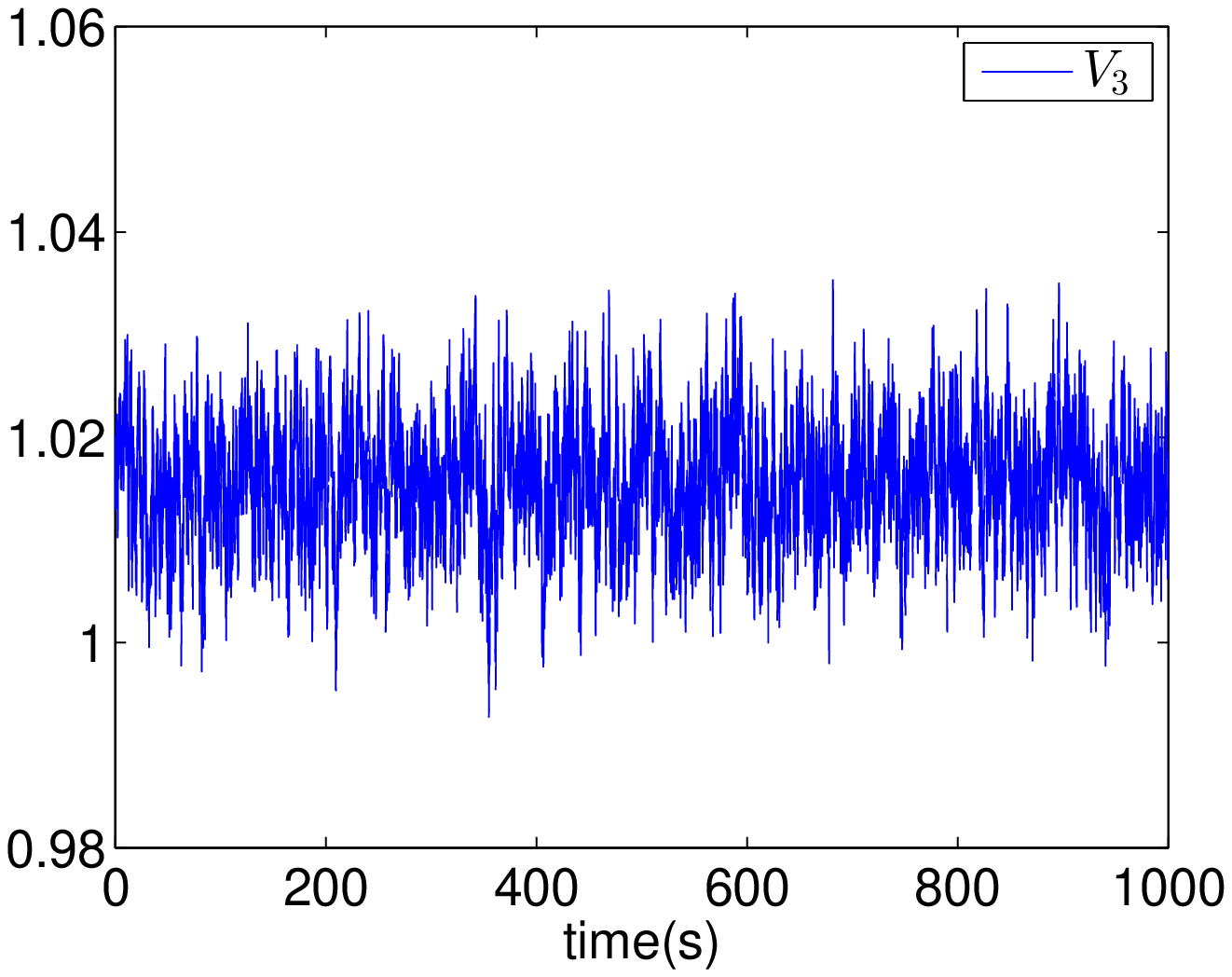}
\caption{Trajectory of $V_3$ on [0s,1000s]}\label{b3}
\end{subfigure}
\caption{Trajectories of some dynamic conductances, susceptances, and voltages in the 9-bus system.}\label{9bus}
\end{figure}

We consider the standard WSCC 3-generator, 9-bus system model (see, e.g. \cite{Chiang:book}). The classical generator models (\ref{swing-1})-(\ref{swing-2}) and the stochastic dynamic load models (\ref{stoloaddynamicp})-(\ref{stoloaddynamicq}) are implemented in the structure preserving framework. The system parameters are available online: https://github.com/xiaozhew/PES-load-parameter-estimation. Particularly, there are three dynamic loads at buses 1, 2 and 3, the time constants of which are $\tau_g=1,3,0.2\mbox{s}$ and $\tau_b=5,7,0.8\mbox{s}$, respectively. The trajectories of some state variables and algebraic variables are shown in Fig. \ref{9bus}, from which we see that the state variables are fluctuating around their nominal values in ambient conditions, yet larger time constants lead to slower variations as expected (e.g., the variations of $g_2$ and $b_2$ are slower than $g_3$ and $b_3$).

By the proposed algorithm, we firstly compute the sample mean $\bar{V}=\mbox{diag}[0.9952, 1.0126,1.0155]$. Then we estimate the dynamic conductance, susceptances and their corresponding sample covariance matrixes:
\begin{eqnarray}
Q_{\bm{gg}}=\left[\begin{array}{ccc}
   1.41\times10^{-3} &  9.84\times10^{-5} &  2.24\times10^{-4}\\
   9.84\times10^{-5} &  4.16\times10^{-4} &  6.57\times10^{-5}\\
   2.24\times10^{-4} &  6.57\times10^{-5}  & 5.75\times10^{-3}
\end{array}\right]
\end{eqnarray}
\begin{eqnarray}
Q_{\bm{bb}}=\left[\begin{array}{ccc}
   2.63\times10^{-4} & -1.47\times10^{-6} &  6.46\times10^{-6}\\
  -1.47\times10^{-6} &  1.75\times10^{-4} & -7.57\times10^{-6}\\
   6.46\times10^{-6} & -7.57\times10^{-6} &  1.62\times10^{-3}
\end{array}\right]
\end{eqnarray}
It is expected that both $Q_{\bm{gg}}$ and $Q_{\bm{bb}}$ are nearly diagonal as the stochastic perturbations are independent.

Since each entry of $(P^s)^2(\Sigma^p)^2$ and $(Q^s)^2(\Sigma^q)^2$ is set to be $0.0025$, $T_g$ and $T_b$ can be readily estimated from (\ref{Tg})-(\ref{Tb}). A comparison between the estimated $\tau_g$, $\tau_b$ and their actual values are shown in Table. \ref{taugb-9bus}. It's observed that the proposed algorithm provides fairly accurate estimation for time constants of each load.

\begin{table}[!ht]
\centering
\caption{A comparison between the actual and the estimated time constants in the 9-bus system}\label{taugb-9bus}
\begin{tabular}{|c|c|c|c|}
\hline
&actual value (s)&estimated value (s) &error\\
\hline
$\tau_{g1}$&1.0000&0.9145&8.55\%\\
\hline
$\tau_{g2}$&3.0000&2.9867&0.44\%\\
\hline
$\tau_{g3}$&0.2000&0.2122&6.1\%\\
\hline
$\tau_{b1}$&5.0000&4.7974&4.05\%\\
\hline
$\tau_{b2}$&7.0000&6.9777&0.32\%\\
\hline
$\tau_{b3}$&0.8000&0.7462&6.72\%\\
\hline
\end{tabular}
\end{table}

\subsection{Impact of Measurement Noise}

Like other measurement-based methods, the performance of the proposed algorithm may be affected by PMU measurement noise. In order to investigate the potential influence, measurement noises with standard deviation of $10^{-3}$ have been added to $\bm{g}$, $\bm{b}$ and ${V}$ in the 9-bus example shown in Section \ref{sectionvalidation} according to the IEEE Standards \cite{IEEEstandard}\cite{IEEEStandard_amd}. A comparison between the actual and the estimated time constants are presented in Table. \ref{taugb-9bus-noise}. It is observed that the proposed method provides similar accuracy to the case without the measurement noises, which indicates that the method is relatively robust under measurement noise.

\begin{table}[!ht]
\centering
\caption{A comparison between the actual and the estimated time constants in the 9-bus system with the measurement noises}\label{taugb-9bus-noise}
\begin{tabular}{|c|c|c|c|}
\hline
&actual value (s)&estimated value (s) &error\\
\hline
$\tau_{g1}$&1.0000&0.9144&8.56\%\\
\hline
$\tau_{g2}$&3.0000&2.9819&0.60\%\\
\hline
$\tau_{g3}$&0.2000&0.2121&6.06\%\\
\hline
$\tau_{b1}$&5.0000&4.7752&4.50\%\\
\hline
$\tau_{b2}$&7.0000&6.9426&0.82\%\\
\hline
$\tau_{b3}$&0.8000&0.7443&6.97\%\\
\hline
\end{tabular}
\end{table}

\subsection{Further Validation}

For further validation, we apply the method to a larger system---the IEEE 39-bus 10-generator test
system, the parameters of which are available online: https://github.com/xiaozhew/PES-load-parameter-estimation. In particular, 10 dynamic loads have been added to buses 1-10, and their corresponding time constants range from $0.1$s to $5$s. A comparison between the actual and the estimated time constants are presented in Table. \ref{taugb-39bus}. The simulation results further demonstrate that the proposed method is able to provide good estimations for time constants of the dynamic loads.

\begin{table}[!ht]
\centering
\caption{A comparison between the actual and the estimated time constants in the 39-bus system}\label{taugb-39bus}
\begin{tabular}{|c|c|c|c|}
\hline
&actual value (s)&estimated value (s) &error\\
\hline
$\tau_{g1}$&0.1000& 0.1225&22.55\%\\
\hline
$\tau_{g2}$&0.6000&0.5860&2.33\%\\
\hline
$\tau_{g3}$&1.1000&1.1086&0.78\%\\
\hline
$\tau_{g4}$&1.6000&1.6924&5.78\%\\
\hline
$\tau_{g5}$&2.1000&2.1323&1.54\%\\
\hline
$\tau_{g6}$&2.6000&2.6130&0.50\%\\
\hline
$\tau_{g7}$&3.1000&3.0715&0.92\%\\
\hline
$\tau_{g8}$&3.6000&3.3004&8.32\%\\
\hline
$\tau_{g9}$&4.1000& 4.5080&9.95\%\\
\hline
$\tau_{g10}$&4.6000&4.6636&1.38\%\\
\hline
$\tau_{b1}$&0.5000&0.5277&5.53\%\\
\hline
$\tau_{b2}$&1.0000&0.9831&1.69\%\\
\hline
$\tau_{b3}$&1.5000&1.5325&2.17\%\\
\hline
$\tau_{b4}$&2.0000&2.0559&2.79\%\\
\hline
$\tau_{b5}$&2.5000&2.5784&3.14\%\\
\hline
$\tau_{b6}$&3.0000&3.3522&11.74\%\\
\hline
$\tau_{b7}$&3.5000&3.1273&10.65\%\\
\hline
$\tau_{b8}$&4.0000&4.1105&2.76\%\\
\hline
$\tau_{b9}$&4.5000&4.3084&4.26\%\\
\hline
$\tau_{b10}$&5.0000&5.1723&3.45\%\\
\hline
\end{tabular}
\end{table}



\section{conclusions and perspectives}\label{sectionconclusion}

In this paper, we have proposed a novel method to estimate parameter values of dynamic load from ambient PMU measurements. The accuracy and robustness of the method have been demonstrated through numerical studies. Unlike conventional methods, the proposed technique does not require the existence of large disturbance to systems, and thus can be implemented continuously in daily operation to provide up-to-date dynamic load parameter values. 

In the future, we plan to further validate the method by using real PMU data and extend the method to estimate dynamic load parameters without knowing their static characteristics.

\end{document}